\documentclass[useAMS]{mn2e}

\usepackage{amssymb,amsmath,psfig,times}
\def\gsim{ \lower .75ex \hbox{$\sim$} \llap{\raise .27ex \hbox{$>$}} }
\def\lsim{ \lower .75ex\hbox{$\sim$} \llap{\raise .27ex \hbox{$<$}} }

\def\sc{Schwarzschild}
\def\beq{\begin{equation}}
\def\eeq{\end{equation}}
\def\msun{$M_\odot$~}
\def\MgII{\hbox{Mg$\scriptstyle\rm II$}}
\def\CIV{\hbox{C$\scriptstyle\rm IV$}}

\title[Jets in high redshift SMBHs]
{The role of relativistic jets in the heaviest and most active supermassive black holes at high redshift}

\author[Ghisellini et al.]
{G. Ghisellini$^1$\thanks{Email:
gabriele.ghisellini@brera.inaf.it}, F. Haardt$^{2,3}$, R. Della Ceca$^{4}$, 
M. Volonteri$^5$, T. Sbarrato$^{1,2}$  \\
$^1$INAF -- Osservatorio Astronomico di Brera, Via Bianchi 46, I--23807 Merate, Italy\\
$^2$DiSAT, Universit\`a degli Studi dell'Insubria, Via Valleggio 11, 22100 Como, Italy\\
$^3$INFN, Sezione di Milano-Bicocca, Piazza delle Scienze 3, 20126, Milano, Italy\\
$^4$INAF -- Osservatorio Astronomico di Brera, Via Brera 28, I--22100 Milano, Italy\\
$^5$IAP, 98bis Boulevard Arago, F-75014 Paris, France\\
}

\begin{document}  

\maketitle

\begin{abstract}
In powerful radio--quiet active galactic nuclei (AGN), the bulk of the population of black holes heavier than
one billion solar masses form at a redshift $\sim$1.5--2.
Supermassive black holes in jetted radio--loud AGN seems to form earlier,
at a redshift close to 4.
The ratio of {\it active} radio--loud to radio--quiet AGN 
hosting heavy black holes is therefore a strong function of redshift.
We report on some recent evidence supporting this conclusion, gathered from
the Burst Alert Telescope (BAT, onboard {\it Swift}) and by the Large Area
Telescope (LAT, onboard {\it Fermi}).
We suggest that the more frequent occurrence of relativistic jets in the most massive black 
holes at high redshifts could be due to the average black hole 
spin being greater in the distant past, or else to the jet helping a fast accretion rate 
(or some combination of the two scenarios).
We emphasize that the large total accretion efficiency of rapidly spinning
black holes inhibits a fast growth, unless a large fraction of the available
gravitational energy of the accreted mass is not converted into radiation, but 
used to form and maintain a powerful jet.

\end{abstract}

\section{Introduction}

How supermassive black holes (SMBH) gained most of their mass is one of the key question 
in modern physical cosmology, and yet there is no general consensus on the kind of evolution 
that such population, as a whole, experienced, particularly at the highest redshifts probed 
but current observations. 

Some SMBHs with masses in excess of $10^9$\msun were already in place 
when the Universe was only $\simeq 700$ Myrs old (e.g., ULAS J1120+0641; Mortlock et al. 2011). 
The very existence of such objects may be difficult to reconcile to black hole growth at the 
Eddington rate starting from stellar sized seeds, and more massive seeds 
may be a  more viable option (see Volonteri 2010 for a review). 
However, such early SMBHs are most probably exceptional objects, rather than the norm. 
Indeed, the global population of very luminous radio--quiet AGNs shows a peak in its number density 
at $z\simeq$2--3, corresponding to a then Hubble time of $\simeq$3--2 Gyrs (see, e.g., 
Hopkins et al. 2007). 
The number of fainter AGNs peaks at even later cosmic time, producing an evolutionary 
luminosity--dependent pattern commonly referred to as ``downsizing". 

Recently, Salvaterra et al. (2012) placed limits on the global accretion history of SMBHs 
at $z\gsim 5$ through the unresolved fraction of the X--ray background 
(see also Dijkstra et al. 2004, Salvaterra et al. 2005, Salvaterra et al. 2007, McQuinn 2012). 
The stacking analysis of the X--ray emission of the $i$--dropouts selected 
by Bouwens et al. (2006) in the {\it Chandra} Deep Field--South provides even 
tighter constrains on the gas accreted onto SMBHs at high redshifts (Willott 2011, 
Fiore et al. 2012, Cowie et al. 2011). 
All those studies concluded that there is some tension between the fast mass accretion 
required by the existence of SMBHs at $z\gsim 6$, and the limit posed by X--ray data. 
Vigorous accretion in the first Gyrs must have proceeded in rare, selected objects, 
and/or in a radiatively inefficient fashion.

The search for the heaviest SMBHs at high-$z$ relies on wide-field optical/IR 
surveys (such as the United Kingdom Infrared Deep Sky Surveys, UKIDSS, or the Sloan 
Digital Sky Survey, SDSS), with subsequent high resolution follow--up spectroscopy from 
8--meter class telescopes. 
Virial--based arguments applied to hydrogen and metal (e.g., \CIV\ and \MgII) emission 
features  then provide an estimate of the black hole mass 
(e.g. McLure \& Dunlop 2004 with H$\beta$ and \MgII; Vestergaard \&
Peterson 2006 with H$\beta$ and \CIV; Vestergaard \& Osmer 2009 with \MgII).
A different, complementary method to probe SMBHs at high $z$ has been proposed by 
Ghisellini et al. (2010a; see also Volonteri et al. 2011; see Calderone
et al. 2013 for a detailed description of the method), who 
showed how high redshift blazars can be used as a proxy of a much larger undetected 
population of radio loud AGNs. 

Blazars are radio loud AGNs whose relativistic jet points directly at us, i.e., our viewing 
angle with respect to the jet axis is $\theta_{\rm v}<1/\Gamma$, where $\Gamma$ is the bulk 
Lorentz factor of the jet.  
Under such circumstances, the non--thermal emission is greatly amplified by relativistic beaming, 
making them well visible also at high cosmic distances. 
Generally speaking, the overall spectral energy distribution (SED) 
of powerful blazars is characterized by two broad distinctive 
humps: a radio/millimeter peak due to primary synchrotron losses, and a second, higher energy 
emission (peaking in the hard X/$\gamma$--rays) resulting from inverse Compton scattering of 
the primary synchrotron photons, and/or of a radiation field produced externally to the jet.
Most powerful blazars, namely flat spectrum radio quasars with optical broad emission lines,
emit most of their luminosity in the high energy hump, that in these sources
is located at 1--100 MeV.
Although this  energy band is difficult to observe, hard X--ray (i.e. above 10 keV) 
and $\gamma$--ray (i.e. above 100 MeV) observation are the most useful to find out
the most powerful jets, because the observational band is close to the peak
of the electromagnetic output of these sources.

\begin{figure}
\vskip -0.5 cm
\vskip -0.1cm
\psfig{figure=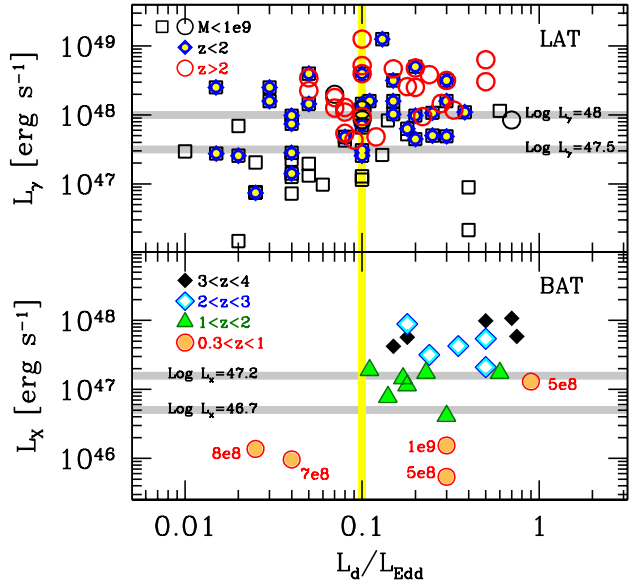,width=9cm,height=9cm}
\vskip -0.5 cm
\caption{
LAT $\gamma$--ray luminosity $L_\gamma$ (top panel) and the 15--55 keV BAT luminosity
$L_X$ vs. the disk luminosity in Eddington units of blazars. 
We show the values derived for all {\it Fermi} blazars at $z>2$,
and for the bright {\it Fermi} blazars detected during the first 3 months of operations. 
For BAT, we show all blazars in the Ajello et al. (2009) 
sample with $z>0.3$. Different symbols correspond to different redshift bins (as labelled).
When not indicated, the black hole mass is above one billion solar masses.
Horizontal grey lines mark  specific threshold luminosities as discussed in the main text. 
}
\label{ll}
\end{figure}

Beaming makes blazars a unique tool in assessing the number density of radio--loud SMBH at high redshift. 
In fact, for any confirmed high--redshift blazar there must exist other 
$2\Gamma^2 =450 (\Gamma/15)^2$ sources sharing the same intrinsic properties, whose jets are 
simply not pointing at us. 
In Ghisellini et al. (2010a) and Volonteri et al. (2011) this idea was explored, and  
we showed that the comoving number density of SMBHs (with $M>10^9M_\odot$)
powering {\it Swift}/BAT (Gehrels et al. 2004) blazars peaks at $z\simeq 4$ 
(see also Ajello et al. 2009).
This was somewhat surprising, since the most luminous radio--quiet quasars 
(i.e. $L_{\rm opt}>10^{47}$ erg s$^{-1}$, thus powered by a black hole with  
$M>10^9 M_\odot$ in order to be sub--Eddington), have a corresponding density 
peaking at lower redshift, $z\sim 2$. 
SMBHs in radio--loud objects seem to be evolved earlier (and possibly faster) 
than their radio quiet counterparts, with a peak of activity 
when the Universe was $\sim$1.5 Gyrs old (to be compared to  $\sim$3 Gyrs for radio--quiet AGNs). 
Such result may have deep consequences for our understanding of the cosmic evolution
of black holes, and of their host galaxies.

However, the fact that the peak of activity of luminous blazars is observed at a 
relatively higher redshift compared to bright radio--quiet AGNs could be merely an artifact due to
the limited bandpass of 
BAT\footnote{15--150 keV, but the study of Ajello et al. (2009) was restricted to 
the 15--55 keV energy band, to have a cleaner signal.}. 
Many more blazars could exist at $z\lsim 4$, simply not detectable 
by BAT because their Compton emission peaks off the hard X--ray band. 
The obvious other band where a relativistic jet can emit most of its electromagnetic 
output is the $\gamma$--rays, probed by, e.g., {\it Fermi}/LAT (Atwood et al. 2009). 
In this paper we employ the recently
published $\gamma$--ray luminosity function (Ajello et al. 2012)
to investigate the existence of a large number of active blazars
(hosting the most massive and active SMBHs) between $1\lsim z \lsim 4$, and discuss the 
possible consequences of our search. 

\section{Heavy and active black holes in blazars}

\subsection{Selection criteria}

In high redshift (powerful) blazars the accretion disk component is
directly visible, as the beamed synchrotron emission contributes mainly at lower frequencies. 
Most of the blazar luminosity is however emitted at even higher frequencies, 
chiefly in the hard X--ray and in the $\gamma$--ray bands. 
Thus, powerful blazars at high redshifts can be found by surveying these spectral windows. 

Two all sky surveys in the hard X--rays (above 15 keV) have been recently carried out,  
by {\it Swift}/BAT (Ajello et al. 2009) and {\it INTEGRAL} (Beckmann et al. 2009), 
while in $\gamma$--rays band the {\it Fermi}/LAT survey (Ajello et al. 2012) is now available. 
Despite the limited sensitivity of all these instruments, it was 
possible to characterize the luminosity functions of blazars in both bands. 
This was done by Ajello et al. (2009) for BAT, and by Ajello et al. (2012) for LAT, and our considerations 
will rely on such two studies. 

Our main goal is to obtain a fair estimate of the number density of the most massive SMBHs accreting 
at high rate, as a function of the cosmic time. 
We therefore identify the following two criteria to select heavy and actively accreting SMBHs:  
\begin{enumerate}
\item $M>10^9 M_\odot$
\item $(L_{\rm d}/ L_{\rm Edd}) >0.1$.
\end{enumerate}
Here $L_{\rm d}$ is the luminosity of the accretion disk, and 
$L_{\rm Edd} \simeq 1.3\times 10^{47}$ $M/(10^9 M_\odot)$ erg s$^{-1}$ is the Eddington luminosity.
Thus, our criteria correspond to disk luminosities in the interval 
$1.3\times 10^{46}< L_{\rm d}<1.3\times 10^{47}$ erg s$^{-1}$. 
The first criterion defines black holes that we dub as ``heavy", while the 
second criterion selects those that are ``active". The two criteria must be met simultaneously by a given source in
order for it to be selected. These two selection criteria will be adopted for both radio--quiet and radio--loud AGNs.
On the other hand, we must consider that blazars, regarded as signposts of the 
entire population of jetted AGNs (see e.g. Ghisellini et al. 2010a), 
are found because of their beamed jet radiation, not on the basis of any accretion 
disk emission (observable only in  the most luminous objects). 
Therefore, we face the need to ``translate" the above requirement on $L_{\rm d}$ into a constraint 
on the non--thermal, hard X--ray through $\gamma$--ray luminosity of such objects. 
To this aim, we rely on our previous studies on BAT and LAT blazars 
(Ghisellini et al. 2009, 2010a, 2010b, 2011). 

In Fig. \ref{ll} the K--corrected rest frame blazars luminosities $L_\gamma$ in the LAT and $L_X$ in the BAT 
bands (upper and lower panel, respectively) are show as a function of $L_{\rm d}/L_{\rm Edd}$.
The reported values for luminosities and masses are derived for all {\it Fermi} blazars above $z=2$
in the 1LAC catalog (Abdo et al. 2010), and for all the bright {\it Fermi} blazars with 
redshift detected during the first 3 months of operations. 
For BAT, we show the values for all blazars in the Ajello et al. (2009) 
sample above $z=0.3$ (colors coded according to different redshift bins as labelled). 

For BAT blazars, a luminosity threshold\footnote{Luminosities are intended in erg s$^{-1}$, 
unless otherwise specified.} 
$\log L_X > 46.7$ includes all heavy and active MBH, at all $z$ larger than $0.3$,
with only one interloper and one outlier. 
Moreover, we note that all the BAT blazars at $z>2$ are powered by heavy and active black holes,
and all have $\log L_X > 47.2$.
We then conclude that a fair, observationally motivated threshold luminosity to select heavy and 
active BAT blazars lies in between the two above mentioned values: $\log L_X > 46.7$ and $\log L_X > 47.2$.

For LAT blazars, a luminosity threshold $\log L_\gamma  > 48$ selects preferentially heavy and active 
blazars as required, most at $z>2$. With this choice we include, however, 
some blazars accreting at less than 10\% Eddington while,
on the other hand, we exclude some objects powered by heavy and active black holes (preferentially at $z<2$).
If we  lower the luminosity divide by a factor of 3 $\log L_\gamma  > 47.5$, we include all the latter
sources, but also many non--active (in our sense) ones. 
By further decreasing the luminosity threshold we would include only lower mass and not so active
blazars. We therefore conclude that the best luminosity threshold for LAT blazars lies in between the two values
marked by the horizontal grey lines:  $\log L_\gamma  > 47.5$ and $\log L_\gamma  > 48$.

\subsection{{\it Swift}/BAT blazars}

Ajello et al. (2009) estimated the hard [15--55 keV] X--ray luminosity function
(LF) of blazars observed by BAT during the first 3 years of the {\it Swift} mission.
Because of its primary scientific goal (i.e., observations of Gamma Ray Bursts), BAT covered
the entire sky was homogeneously, with a sensitivity slightly better than 1 mCrab. 
After classifying all the detected sources (excluding the Galactic plane, $|b| \ge 15^\circ$), 
we can count 38 blazars. Among them, 5 lie between $2<z<3$,  while 5 at $z>3$.

In Ghisellini et al. (2010a), we have studied all the 10 BAT 
blazars\footnote{We have also studied the BAT blazars with $0.3<z<2$ shown in Fig. \ref{ll}. 
Details about the model results will be presented elsewhere.}
at $z>2$, finding that all of them host a black hole with a mass exceeding 
$10^9 M_\odot$, and all emit $\log L_X  > 47.2$. 
The associated accretion disks produce an average luminosity 
$\langle L_{\rm d}/L_{\rm Edd}\rangle \sim 0.3$ (see Fig. 1 in Volonteri et al. 2011). 
On these basis, we could calculate the expected density of blazars powered by active and heavy black holes
as a function of $z$, integrating the LF above $\log L_X =46.7$, 
and above $\log L_X =47.2$. 

Some extra care in estimating the LF at  $z\sim 4$ was needed, since no blazars was detected by BAT 
beyond such redshift. At variance with Ajello et al. (2009), we then decided to introduce 
an exponential cut--off on the luminosity evolution of blazars above $z=4$, in order to derive 
a number density at the same time conservative (predicting the
minimum number of blazars) and consistent with the very few blazars observed by other 
X--ray satellites at $z>4$  (see Fig. 16 in Ghisellini et al. 2010a).
The resulting expected density of blazars powered by heavy and active black holes is 
shown in Fig. \ref{fi2} as a red stripe.


{\subsection{{\it Fermi}/LAT blazars}

Also {\it Fermi}/LAT detected several blazars at $z > 2$.
There are 31 blazars with $z>2$ in the 2LAC catalog (Ackermann et al. 2011),
which refers to the first 2 years of operations. The redshift distributions of blazars 
resulting from the two surveys (BAT and LAT) are rather 
different since, despite the fact that the total number of LAT blazars is substantially larger than
the one of BAT,  only a couple were detected by {\it Fermi} at $z\gsim3$.

In our previous works (Ghisellini et al. 2009; 2010b; 2011)
we have shown that all LAT blazars with $\log{L_\gamma}> 48$ host black holes with masses  
$\gsim 10^9 M_\odot$ accreting at relatively large rates, $\langle L_{\rm d}/L_{\rm Edd}\rangle \sim 0.1$, 
although, on average, smaller than the $z>2$ BAT blazars ($\langle L_{\rm d}/L_{\rm Edd}\rangle \sim 0.3$). 

Ajello et al. (2012) derived the luminosity function of blazars detected  by {\it Fermi}--LAT. 
Overall, the $\gamma-$ray LF evolves quite rapidly up to
$z \simeq 1.5-2$, and declines at larger redshifts. 
Integrating the LF above
$\log L_\gamma=47.5$ ($\log L_\gamma>48$) gives the the number density of heavy and active LAT blazars shown as 
the upper (lower) green curve in Fig. \ref{fi2}.

\section{Results}

Fig. \ref{fi2} encompasses several important pieces of information about 
the evolution of heavy and active black holes. 
The comoving number density of jetted sources matching our selection criteria 
($M>10^9M_\odot$ and $L_d/L_{\rm Edd}>0.1$) is shown as a function of redshift. 
We assumed that for each {\it observed} blazar there are 450 other misaligned sources not pointing at us, and thus undetectable. 
This means that we assumed an average value of 15 for the bulk Lorentz factor of the jet in 
such objects, consistent with the results of the extensive SED model fitting procedure 
performed by Ghisellini et al. (2010b). 

The redshift distribution of LAT blazars (green curves and stripe) seems to peak at relatively 
later epoch, $z\simeq1-2$, while BAT blazars (red curves and stripe)  show a 
different behavior, with a  peak at $z\simeq 4$ (the sharpness of the peak may be an artifact of our 
conservative approach that introduced an exponential cut-off at $z>4$).
The combined redshift distribution reveals that the number density of blazars still 
peaks at a high redshift, $z\simeq 4$, or alternatively stays approximatively  constant 
in the range $1\lsim z \lsim4$ if one takes $\log L_X=47.2$ and $\log L_\gamma=47.5$ 
as luminosity thresholds for the BAT and LAT samples, respectively. 
Note how the density of BAT blazars changes more (by decreasing the limiting $L_X$) than the 
density of LAT blazars. 
Whatever the limiting luminosities are, Fig. \ref{fi2} clearly shows that LAT blazars are in any case 
not numerous enough to shift the peak of black hole main activity at more recent epochs. 
We conclude that the heaviest black holes with jets formed by $z\gsim 4$.

We then compared the number density of blazars with the 
number density of powerful radio--quiet QSOs. 
To this aim, we integrated the bolometric LF derived from the SDSS survey by Hopkins et al. (2007), 
above three minimum luminosities, namely $\log L_{\rm bol}=46$, 46.5, and 47, as 
shown in Fig. \ref{fi2} (blue curves, from top to bottom). 
We note that, according to Richards et al. (2006), the bolometric luminosity is approximatively 
twice as large as the accretion disk luminosity (see also Calderone et al. 2013). 
Correspondingly, the three luminosity thresholds translate into 
$\log L_{\rm d}=45.7$, 46.2, and 46.7. For a black hole of mass $M=10^9 M_\odot$, 
the corresponding three Eddington ratios are then 0.04, 0.12 and 0.4.
Note that such limits are less tight than those we applied to X--ray and $\gamma-$ray samples, 
since we allow for Eddington ratios smaller than 0.1, and/or black holes lighter than $10^9 M_\odot$. 
Our goal here is, in fact, a conservative assessment of the relative occurrence of jetted AGNs 
compared to the entire AGN population. 

Results of the integration of the Hopkins et al. (2007) LF show that that the number density of 
powerful radio--quiet QSOs peaks at $z\simeq 2-$2.5. 
By decreasing the luminosity threshold, the number 
density increases, and the peak of the redshift distribution shifts to slightly smaller redshifts. 
The evidence we can gather from our combined analysis of 
hard X--rays, $\gamma$--rays and optical AGN surveys suggests that two different
epochs for the growth of SMBHs exist, according to the presence/absence of relativistic jets.
If the jet is present, the heaviest black holes grow by $z\simeq 4$, while the growth of heavy 
non--jetted black holes is delayed at $z\simeq$2.

The grey stripe in Fig. \ref{fi2} indicates the number density of haloes exceeding a
given threshold mass, defined as the minimum halo mass to host a $M>10^9 M_\odot$ black hole
(Ghisellini et al. 2010a). 
This can be regarded as a fiducial upper limit to the possible density of 
$10^9 M_\odot$ black holes at high redshifts. 
As already discussed, even in the limiting case of luminosity thresholds  
$\log L_X > 47.2$ and $\log L_\gamma > 47.5$, the redshift distribution of 
the heavy black holes in jetted--AGNs would be quasi--flat, very different
from radio--quiet QSOs. 
This very fact bears the important consequence that the radio--loud fraction of powerful 
AGNs is a strong function of redshift or, in other words, that heavy and active black holes 
are more and more associated with jets as the redshift increases.

\begin{figure}
\vskip -0.6 cm
\hskip -0.4 cm
\psfig{figure=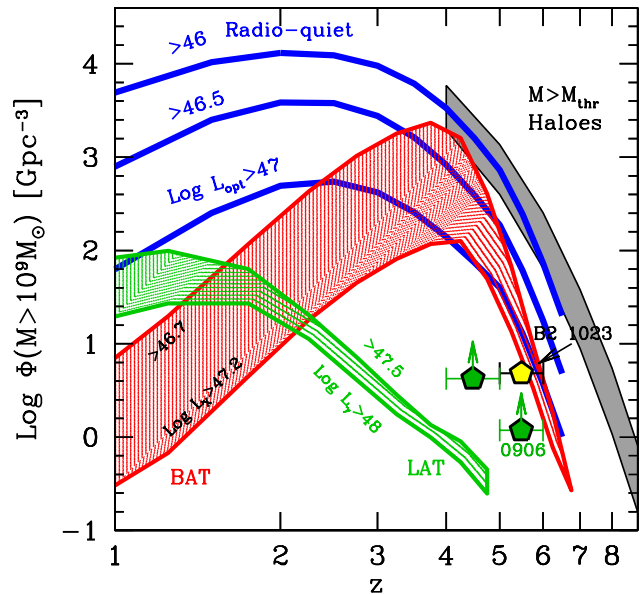,width=9.5cm,height=9.5cm}
\vskip -0.9 cm
\caption{
Comoving number density of  blazars powered by ``heavy and active" black holes 
($M>10^9M_\odot$, $L_{\rm d}/L_{\rm Edd}>0.1$) as a function of redshift.
The larger (red in the electronic edition) hatched band is derived by integrating the [15--55 keV]
luminosity function (Ajello et al. 2009, modified as in Ghisellini et al. 2010a)
above 
$\log L_X =46.7$ (upper boundary) and
$\log L_X =47.2$ (lower boundary), and multiplying
the derived density by 450 (i.e. $2\Gamma^2$, with  $\Gamma=15$).
The smaller (green) hatched band is derived by integrating the 
$\gamma$--ray luminosity function above 
$\log L_\gamma=47.5$ (upper boundary) and
$\log L_\gamma =48$ (lower boundary). 
All sources selected should own a disk with $L_{\rm d}>0.1L_{\rm Edd}$, 
accreting onto a SMBH of mass $>10^9 M_\odot$.  
The tree (blue) stripes are derived integrating the luminosity function of
radio--quiet quasars (Hopkins et al. 2007), above three different threshold
luminosities, as labelled.
The grey stripe is based on connecting black hole mass to halo 
mass, as described in \S 7.2 of Ghisellini et al. (2010a); $M_{\rm thr}$
is the minimum halo mass required to host a $10^9 M_\odot$ black hole. 
This can be considered as the upper limit to the density of $10^9 M_\odot$ black holes.
The (green) pentagons correspond to the density inferred from the few 
sources at high--$z$ already identified as blazars.
The (yellow) pentagon labelled B2 1023 is the density inferred on the existence of only one blazar,
B2 1023+25, in the region of the sky covered by the SDSS+FIRST surveys
(Sbarrato et al. 2012b).
}
\label{fi2}
\end{figure}

\section{Discussion}

The main result of our study is that the most massive black holes in the high redshift Universe 
were more likely associated with relativistic jets. 
Next we will discuss whether this implies the preference for jets to be 
created in more massive systems, or if rather supports the opposite, namely that 
the presence of a jet helps the black hole to grow faster. We have also shown that the 
chance for an active and heavy black hole to launch a jet is greatly reduced at later 
times ($z\sim 2$), i.e., the AGN is more likely quiet in the radio band. 


\subsection{Accretion driven spin--up}

The common presence of jetted SMBHs at high redshift qualitatively fits in the global model for 
mass and spin evolution recently presented by Volonteri et al. (2012). 

Sustained accretion  grows the most massive SMBHs at early times, i.e., 
the mass of the most active SMBHs is larger at high redshifts ($z\gsim 4$). 
The result of the massive, coherent accretion triggered by gas--rich major mergers 
is a rapid spin--up of the hole, with a spin parameter always close to maximum for the 
most active, most massive SMBHs powering quasars. 
{\it If} these high spins are required to launch jets, then there should be plenty of them at 
early epochs, including jets associated with the most massive black holes. 
By $z\lsim 2$, instead, the most massive galaxies become poorer in gas.  
This has two effects. 
One, SMBH--SMBH mergers occurring in a gas--poor galaxy are not expected 
to have any preferential symmetry or alignment (Bogdanovic et al. 2007), 
and such mergers lead to a typical spin value of 0.6--0.7 (Berti \& Volonteri 2008). 
Two, in gas--poor galaxies lacking central nuclear discs, accretion occurs in a more chaotic 
fashion, making the distribution of spins less skewed towards high values. 
This shallower distribution translates into a reduced ratio of 
radio--loud to radio--quiet objects, including the 
sources with the most massive black holes. 

\subsection{High accretion rates in jetted disks}

In an alternative scenario the jet itself is a key cause determining the global accretion 
properties of the system, rather than a mere effect of massive coherent accretion, as discussed above. 

In a radiatively efficient accretion flow the gravitational energy of the matter accreting 
at a rate $\dot M$ is transformed into heat, and then into radiation. 
The disk luminosity is $L_{\rm d}=\eta_{\rm d} \dot M c^2$,  
where $\eta_{d}\approx 0.1$ is the efficiency of converting mass into radiation. 
Jolley \& Kuncic (2008) and Jolley et al. (2009) suggested  that part of
the dissipation of gravitational energy can lead to 
amplification of magnetic field in the disk, rather than to heat production. 
The amplified magnetic field may drive different phenomena observed in AGNs, specifically the 
formation of i) a X--ray emitting, hot corona, ii) a fast, yet sub--relativistic
outflow, iii) a genuinely relativistic jet (e.g. Sikora \& Begelman 2013). 
Therefore a fraction of the available gravitational energy can be transformed in mechanical 
energy, besides radiation.
As a result, the accreting matter attains a smaller local temperature compared to standard 
disk models, because less energy is dissipated through radiation.
It is then possible that radiative losses in the form of thermal optical--UV emission 
are comparatively modest, while a sizeable fraction of the available gravitational energy goes 
instead to power the corona and/or an outflow. 

In this picture, radio--loud AGNs can share with radio--quiet AGNs hot X--ray coronae and fast 
(yet sub--relativistic) outflows, but in addition the accretion--driven magnetic field can 
be used to directly launch a jet, or for allowing the
extraction of the rotational energy from the black hole.
In other words: the use of gravitational energy to amplify disk magnetic fields 
is not necessarily {\it alternative}
to the Blandford \& Znajek (1977) mechanism, since also in this process
a relatively large magnetic field is required to tap the hole spin energy.
Jets could receive their power both from the accretion and from the black hole spin.

The partial transformation of the gravitational energy into magnetic fields
that help to launch the jet is not accompanied by a significant amount
of mass that is accelerated: 
in other words, 
a fraction of the gravitational energy is transported outwards by a small amount of mass. 
In this case, the efficiency $\eta_{\rm d}$ defining the disk luminosity is smaller
than the standard value and the Eddington limit is reached for larger $\dot M$. 

One can define a global efficiency $\eta$ that is the sum
of two terms, following Jolley \& Kunzic (2008; 
see also Shankar et al. 2008), one for the generated magnetic field, 
and the other for the disk luminosity:
\begin{equation}
\eta \, \equiv \, \eta_{\rm B} +\eta_{\rm d}.
\end{equation}
Disks can be maintained (relatively) cold through the non--thermal dissipation 
of the gravitational energy mediated by the amplification of magnetic fields.
These magnetized disks can sustain large accretion rates before becoming
super--Eddington in view of their reduced temperature. 
As a clear consequence, the black hole grows faster than its radio--quiet counterpart, 
doubling its mass on a  Salpeter (1964) time, which is now:
\begin{eqnarray}
t_{S} \, &=&\, {M\over \dot M} \, =\,
{\eta_{\rm d} \over 1-\eta } \, {\sigma_{\rm T} c \over 4\pi G m_{\rm p}}
\, {L_{\rm Edd}\over L_{\rm d} }
\nonumber \\
 &\simeq& 450 \, \left({ \eta_{\rm d} \over  1-\eta }\right)
\left({ L_{\rm Edd} \over L_{\rm d} }\right)\,\,\, {\rm Myr}.
\label{ts}
\end{eqnarray}
The time $t_9$ needed to form a $10^9 M_\odot$ black hole accreting always
at the Eddington limit, starting from a seed of $10^2 M_\odot$,  is
\begin{equation}
t_9 \,  \sim  \, 806 \,\, { \eta_{\rm d}/0.1 \over (1-\eta)/ 0.9} \,\,\,\,\,  {\rm Myr}
\label{t9}
\end{equation}
If accretion starts (ideally) at infinite redshift, 
then $t_9=806$ Myr corresponds to $z=6.6$ in a standard $\Lambda$CDM cosmology.
%
%
If the accretion process starts at $z = 20$ (when the Universe was 0.18 Gyr old) 
and $\eta=\eta_{\rm d}=0.1$, 
then a SMBH reaches one billion solar masses at the redshift corresponding to 
an age of the Universe of (0.18 Gyr $+ t_9$), corresponding to $z = 5.7$. 
This can be read--off in Fig. \ref{salpeter} following the lower dotÐ-dashed line. 
If we assume the same initial mass and redshift, but a total efficiency $\eta = 0.1$, equally shared 
between $\eta_{\rm B}=\eta_{\rm d}=0.05$, then $t_9$ halves and we obtain (0.18 Gyr $+ t_9$)=0.4 Gyr, 
corresponding to $z = 8.4$. The same black hole grows much faster.

\begin{figure}
\vskip -0.3 cm
\hskip -0.5 cm
\psfig{figure=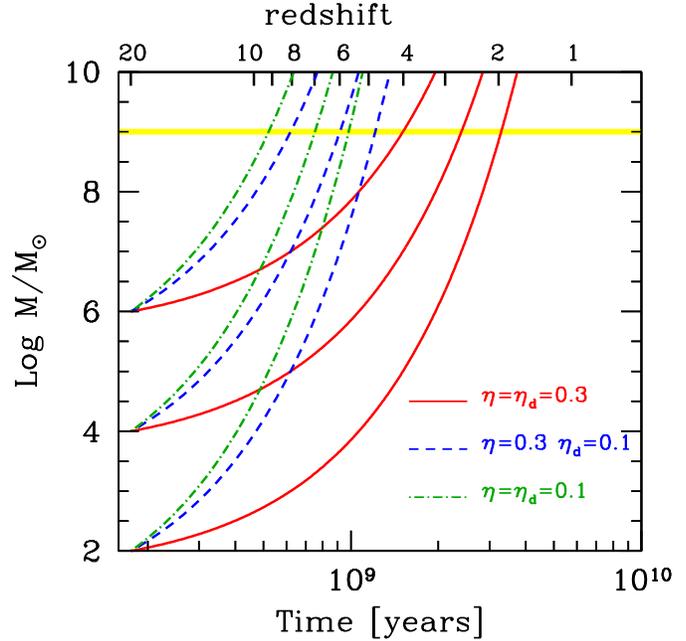,width=9.5cm,height=9.5cm}
\vskip -0.4 cm
\caption{ 
The mass of a black hole accreting at the Eddington rate as
a function of time (bottom axis) and redshift (top axis). 
Accretion starts at $z=20$ onto a black hole seed of $10^2M_\odot$,
 $10^4M_\odot$ or $10^6M_\odot$ , with different efficiencies, as labeled.
The horizontal line marks one billion solar masses.
The larger $\eta_{\rm d}$, the smaller the amount of accreted mass
needed to produce a given luminosity, and the longer the black hole 
growing time. If part of the accretion energy goes into launching a jet,
however,  $\eta_{\rm d}<\eta$ and the growth time decreases. 
}
\label{salpeter}
\end{figure}

\subsubsection{Jets and spinning black holes}

The presence of jets may be associated to rapidly spinning black holes,
since their rotational energy can be an important source of power for the jet
(Blandford \& Znajek 1977; Tchekhovskoy, Narayan\& McKinney 2011)
and the angular momentum of the hole naturally identifies the jet direction.
In fact, it has been proposed that the dichotomy between radio--loud and quiet AGNs might be 
associated to the spin value (e.g. Wilson \& Colbert 1995; Sikora, Stawarz \& Lasota 2007).

Rapidly spinning black holes are characterized by high accretion efficiencies, 
since the innermost stable orbit
for prograde accretion approaches the gravitational radius $R_{\rm g} \equiv GM/c^2$.
Thorne (1974) pointed out that, through accretion, the black hole achieves a maximum
(dimensionless) spin $a=0.998$. 
It fails to achieve the maximum value $a=1$ because of the counteracting torque
of the accretion disk radiation captured by the black hole.
With $a=0.998$, the total efficiency is $\eta=0.3$ (Thorne 1974).
Having a larger $\eta$, rapidly rotating black holes can thus be more efficient in 
producing radiation with respect to non--spinning holes.  
As $\eta_{\rm d}$ increases, a given disk luminosity corresponds to a lower overall 
accretion rate, $\dot M = L_{\rm d}/ (\eta_{\rm d} c^2)$. 
For a fixed luminosity, therefore, a black hole accretes {\it less} matter and therefore 
it will need more time to grow, according to Eq. \ref{ts}.
If $\eta_d= 0.3$ (and $\eta_{\rm B}=0$) 
from the start of the accretion process (set at $z = 20$) then (0.18 Gyr$ + t_9$)=3.3 Gyr, 
corresponding to $z = 2$. 
However, if $\eta_{\rm B}>0$ the growth time decreases; 
for instance halving $\eta_{\rm d}$ by setting $\eta_{\rm B} = 0.15$, 
one obtains ($t_9+0.18$ Gyr)=1.55 Gyr, corresponding to $z=3.6$.

The most distant blazar with $M\gsim 10^9M_\odot$ has $z=5.47$
(corresponding to $\sim$1 Gyr).
If this black hole is maximally spinning, and thus 
$\eta=0.3$, its presence {\it requires} 
$\eta_{\rm B}\ge 0.2$ and $\eta_{\rm d}\le 0.1$.
This is illustrated in Fig. \ref{salpeter} by the lower dashed
(blue) line, corresponding to a black hole seed of 100 solar masses.
Increasing the mass of the seeds helps, but only logarithmically 
(see the middle and top series of lines for a seed of $10^4 
M_\odot$ or $10^6 M_\odot$ respectively).
Even in these cases we are forced to assume that if the black hole
is spinning rapidly, most of the 
liberated gravitational energy is not transformed into radiation,
but into magnetic fields.


The simple model sketched here can, at least qualitatively, account for the results we have 
discussed in the previous section. 
When selecting heavy and active AGNs at high redshift, we introduce a bias towards objects 
where the presence of a jet helped a rapid, vigorous mass accretion, in a gas--rich galaxy. 
We then expect most of the heavy and active black holes at  $z\gsim 4$ to be observed as 
radio--loud sources. 

On the contrary,
at lower redshift galaxies have less cold gas available for black hole accretion.
Still, the possible presence of the jet reduces the disk
efficiency $\eta_{\rm d}$, but this does not translate in a larger $\dot M$, simply because the 
gas reservoir is limited. 
As a consequence, black holes at $z\lsim 2$ do not grow faster even when they have a jet,
and the relative density of radio--loud to radio--quiet AGN with heavy black holes is
smaller than at $z\sim4$.


\section*{Acknowledgments}

We would like to thank the referee, A. Fabian, for useful suggestions
that helped to clarify the paper.
RDC acknowledge financial support from
ASI grant No. I/088/06/0.
MV acknowledges funding support from NASA, through award ATP NNX10AC84G, 
and from a Marie Curie Career Integration grant (PCIG10-GA-2011-303609). 


\end{document}